\documentclass[aps,prb,showpacs,preprint,peprintnumbers]{revtex4}%
\usepackage{amssymb}
\usepackage{amsfonts}
\usepackage{amsmath}
\usepackage{graphicx}
\usepackage{dcolumn}
\usepackage{bm}
\usepackage{textcomp}%
\begin{document}
\title{Spin-Orbit Engineering of Semiconductor Heterostructures }
\author{Federico Bottegoni$^{a}$}
\author{Henri-Jean Drouhin$^{a}$}
\email{Henri-Jean.Drouhin@polytechnique.edu}
\author{Guy Fishman$^{b}$}
\author{Jean-Eric Wegrowe$^{a}$}
\affiliation{$^{a}$ Ecole Polytechnique, LSI, CNRS and CEA/DSM/IRAMIS, Palaiseau, ~F-91128}
\affiliation{$^{b}$Univ Paris-Sud, IEF, CNRS,~Orsay, F-91405}
\date{\today }

\begin{abstract}
We present a systematic construction of the probability-current operator,
based on a momentum power expansion of effective Hamiltonians. The result is
valid in the presence of a Rashba term and when a D'yakonov--Perel
contribution is included. We propose practical tools for spin-orbit
engineering of semiconductor heterostructures. We apply this formalism to a
paradigmatic system, the interface between two semi-infinite media, on one
side a free-electron-like material and on the other side a barrier material
with spin-orbit interaction. We show that the usual boundary conditions,
namely the continuity of the envelope function and of a velocity at the
interface, according to the BenDaniel-Duke approach, comply with the
conservation of the probability current only when first- (Rashba-like) and
second-order (free-electron-like) terms are taken into account in the
Hamiltonian. We revisit the boundary conditions and we prove that the envelope
function may be discontinuous at the interface.

\end{abstract}

\pacs{72.25.Dc, 71.70.Ej, 73.40.Gk}
\maketitle

\section{Introduction}

The probability current is a fundamental concept in quantum mechanics, which
connects the wave-like description of a quasi-particle to the notion of
transport current. When we consider a general Schr\"{o}dinger problem where
the Hamiltonian is%
\begin{equation}
\widehat{H}_{0}=\frac{\widehat{\mathbf{p}}^{2}}{2m}+\mathfrak{U}\left(
\mathbf{r}\right)  \text{,}\label{H0}%
\end{equation}
where the real potential $\mathfrak{U}\left(  \mathbf{r}\right)  $ is periodic
in a crystalline solid and $m$ is the free-electron mass, we are led to the
usual definition of free-electron current probability:\cite{Cohen}%
\begin{equation}
\mathbf{J}^{f}[\psi]=\operatorname{Re}\left[  \psi^{\ast}\frac{\hat
{\mathbf{p}}}{m}\psi\right]  =\frac{\hbar}{m}\operatorname{Im}\left[
\psi^{\ast}\mathbf{\nabla}\psi\right]  \text{.}\label{1}%
\end{equation}

However, in condensed-matter systems in the presence of Spin-Orbit Interaction
(SOI), the potential is no longer real so that a redefinition of this quantity
is mandatory. A debated example of this subtle point is provided by
semiconductors-based systems, whose proper treatment requires consideration of
the Hamiltonian
\begin{equation}
\widehat{H}=\widehat{H}_{0}+\widehat{H}_{SO}\label{H}%
\end{equation}
with%
\begin{equation}
\widehat{H}_{SO}=\frac{\hbar}{4m^{2}c^{2}}\left(  \mathbf{\nabla}%
\mathfrak{U}\times\hat{\mathbf{p}}\right)  \cdot\hat{\mathbf{\sigma}}%
\text{.}\label{2}%
\end{equation}

Following the arguments developed by Hoai Nguyen \textit{et al}.,
\cite{Hoai09} it is reasonable to express the full Hamiltonian, involving SOI
terms, as an effective Hamiltonian which consists of momentum-operator
$\hat{\mathbf{p}}$-power series expansion: Indeed, beside the kinetic energy,
quadratic in $\hat{\mathbf{p}}$, the SOI provides leading terms that are
linear and cubic in $\hat{\mathbf{p}}$, known respectively as Rashba
\cite{Rashba91} and D'yakonov-Perel (DP)\cite{Perel71} terms. Then, since the
SOI potential is non real, a more general definition of the probability
current $\mathbf{J}[\psi]$ has necessarily to be taken into account.
Considering interactions that include higher-order polynomial terms in the
Hamiltonian, we have to deal with an effective Hamiltonian of order $n$.

Furthermore, an open question, strictly linked to the one above, concerns the
definition of spin current (SC). Indeed, in semiconductor physics, that
provides paradigmatic systems for spintronics, it is known that the SC
standard definition, used by many
authors,\cite{Rashba03,Sinova04,Son07,Sabli07,Litvi10,Haney10} can be suitably
applied to two dimensional (2D) systems with Rashba SOI, but fails to describe
spin-dependent transport phenomena in bulk cubic semiconductors, where SOI
induces a DP term in the conduction band. The existence of extra-current terms
was pointed out in Ref.
\onlinecite{LiTao}%
. Drouhin \textit{et al}. \cite{Drouhin11} have shown that a redefinition of
SC is mandatory to obtain a unified treatment, enlightening the fact that a
properly-symmetrized spin-current operator $\hat{\mathbf{J}}_{\uparrow\left(
\downarrow\right)  }[\psi]$, where $\uparrow\left(  \downarrow\right)  $
refers to up (down) spin channel, gives unexpected results when applied to
tunneling through evanescent states in GaAs barriers.

As pointed out by Rashba in Ref.
\onlinecite{Rashba03}%
, there are still concerns relying on the fact that a consistent theory of
spin transport currents has not been formulated yet. From a general point of
view, it means that we cannot immediately approach such a topic in terms of
non equilibrium thermodynamics. In fact, a difficulty relies on the definition
of system in order to formulate relevant balance equations and also on the
boundary terms which should possibly be included in the effective Hamiltonian.
Recently, Shi \textit{et al.} \cite{Shi06} have proposed an alternative
spin-current operator, satisfying the continuity equation, that allegedly
supports important conclusions concerning conservation of spin
currents,\cite{Sugimoto06,Zhang08,Wong10} but which appears to rely on
non-explicit assumptions (see Sec. \ref{sec Gen}).

The inclusion of SOI in the Hamiltonian of a system has direct and practical
consequences in heterostructures, where a consistent analysis of the tunneling
phenomena is required. The pragmatic BenDaniel-Duke (BDD)
approach,\cite{BenDaniel} that perfectly works when dealing with quadratic
Hamiltonians under effective-mass approximation, cannot be straightforwardly
extended because it is not always possible to ensure both the continuity of
the envelope function and the conservation of the probability current, which
is mandatory under steady-state conditions. Then it is necessary to revisit
both the probability-current expression and the boundary conditions. This is
in line with the ideas of Harrison.\cite{Harrison}

In this paper, we present a systematic construction of the probability-current
operator $\hat{\mathbf{J}}$, based on an effective Hamiltonian written as a
$\hat{\mathbf{p}}$-power series expansion. We show the relation between the
velocity operator and the current operator, evidencing the simple structure of
the extra terms. This yields easy and compact calculations whereas explicit
treatments in particular cases resulted in lengthy calculations.\cite{LiTao}
The current operator can be subsequently used to build the SC operators
according to the procedure described in Ref.
\onlinecite{Drouhin11}%
. Then, we introduce proper matching conditions at the boundaries, which
generalize the BDD procedure, the simplest efficient way to deal with
semiconductor heterostructures. Finally, we illustrate our method on three
examples: the case of a quadratic Hamiltonian, where we recover the usual
situation (continuity of the envelope function and of the velocity), the case
where a Rashba term is added as a perturbation to the BDD Hamiltonian (there
we find that the envelope function is continuous, but its derivative is
discontinuous), and the case where a cubic DP term is added to the BDD
Hamiltonian (where we prove that the envelope function cannot be continuous).

The layout of this paper is as follows: Sec. \ref{sec Gen}, we give a general
construction of current operators and a derivation of local properties. Sec.
III, we introduce a general Hamiltonian $\hat{H}^{(n)}$ as a $n^{th}$-degree
homogenous function of momentum-operator coordinates; we consistently derive
the velocity operator and we show that a proper symmetrization yields the
Hermitian current operator $\widehat{\mathbf{J}}$. Sec. \ref{BDD}, we propose
boundary conditions which are suitable to deal with heterostructures. Sec.
\ref{sec GaAs}, we apply our method to electron tunneling through a
$[110]$-oriented GaAs barrier. Sec. \ref{Spin Cur}, we show how to extend the
construction procedure to the spin current operators.

\section{General definition of current operators\label{sec Gen}}

A difficulty, that arises when a current operator is taken into account,
relies on the correct definition of the system and of its boundaries: in fact,
considering the density $\rho$ of a physical quantity, we need to satisfy the
continuity equation of $\mathbf{J}$, defining a source term $G$, so that:%
\begin{equation}
\frac{\partial\rho}{\partial t}+\mathbf{\nabla\cdot J}=G\text{.}\label{3}%
\end{equation}

As pointed out by Shi et al. in Ref.
\onlinecite{Shi06}%
, the continuity of $\mathbf{J}$ can be ensured by introducing a general
source term $G$, as in Eq. \ref{3}, but the source term is not uniquely
defined and this leads to possible confusion when considering the conservation
laws in terms of non-equilibrium thermodynamic equations.\cite{JeanEric} In
any case, we first need to state clearly the local properties of a current
operator, postponing the analysis of its global properties. For this purpose,
we consider a linear operator $\widehat{A}$ that does not explicitly depend on
time and acts over a generic state $\psi$. In the following we adopt the
notation $\left(  \widehat{A}\right)  =\left(  \psi\left\vert \widehat
{A}\,\psi\right.  \right)  =\psi^{\dagger}\widehat{A}\,\psi$ used in Ref.
\onlinecite{Hoai09}%
. The general Schr\"{o}dinger problem reads:%
\begin{equation}
i\hbar\frac{\partial}{\partial t}\psi=\widehat{\mathcal{H}}\,\psi\label{4}%
\end{equation}
where $\widehat{\mathcal{H}}$ may be any Hamiltonian. For example
$\widehat{\mathcal{H}}$ may be equal to $\widehat{H}$ (defined in Eq. \ref{H})
or to $\widehat{H}_{eff}$ (defined below in Eq. \ref{a 37}). We explicitly
develop the derivative of $\widehat{A}$ with respect to time:%
\begin{equation}
\frac{\partial}{\partial t}\left(  \widehat{A}\right)  =\frac{\partial
}{\partial t}\left(  \psi^{\dagger}\widehat{A}\,\psi\right)  =\frac{\partial
}{\partial t}\left(  \psi^{\dagger}\right)  \widehat{A}\,\psi+\psi^{\dagger
}\widehat{A}\left(  \frac{\partial}{\partial t}\psi\right) \label{5}%
\end{equation}
and with the help of Eq. \ref{4} we obtain:%
\begin{equation}
\frac{\partial}{\partial t}\left(  \widehat{A}\right)  =-\frac{1}{i\hbar
}\left(  \widehat{\mathcal{H}}\,\psi\right)  ^{\dagger}\widehat{A}\,\psi
+\frac{1}{i\hbar}\psi^{\dagger}\widehat{A}\left(  \widehat{\mathcal{H}}%
\,\psi\right)  =\frac{1}{i\hbar}\left[  \psi^{\dagger}\,\widehat{A}%
\,\widehat{\mathcal{H}}\,\psi-\left(  \widehat{\mathcal{H}}\,\psi\right)
^{\dagger}\widehat{A}\psi\right]  \text{.}\label{6}%
\end{equation}

If $\widehat{A}$ is an Hermitian \textit{matrix} (the elements of which are
complex numbers, not differential operators)%
\begin{equation}
\left(  \widehat{\mathcal{H}}\,\psi\right)  ^{\dagger}\widehat{A}%
\,\psi=\left(  \psi^{\dagger}\,\widehat{A}\,\widehat{\mathcal{H}}%
\,\psi\right)  ^{\ast}\text{,}\label{80}%
\end{equation}
so that we can rewrite Eq. \ref{6} in a more suitable way that we refer to as
the local form of Ehrenfest theorem:%
\begin{equation}
\frac{\partial}{\partial t}\left(  \widehat{A}\right)  =\frac{2}{\hbar
}\operatorname{Im}\left(  \psi^{\dagger}\,\widehat{A}\,\widehat{\mathcal{H}%
}\,\psi\right)  \text{.}\label{8}%
\end{equation}

The integration over the whole space leads to the well known Ehrenfest's
theorem, whose global form is valid for any (possibly differential) Hermitian
operator $\widehat{A}$:%
\begin{equation}
\frac{\partial}{\partial t}\left\langle \widehat{A}\right\rangle =\frac
{1}{i\hbar}\left[  \left\langle \psi\left\vert \widehat{A}\,\widehat
{\mathcal{H}}\right\vert \psi\right\rangle -\left\langle \widehat{\mathcal{H}%
}\psi\left\vert \widehat{A}\right\vert \psi\right\rangle \right]  =\frac
{1}{i\hbar}\left\langle \psi\left\vert \left[  \widehat{A},\widehat
{\mathcal{H}}\right]  \right\vert \psi\right\rangle \text{.}\label{12}%
\end{equation}

We can write%
\begin{equation}
\frac{\partial}{\partial t}\left(  \widehat{A}\right)  =\frac{1}{\hbar
}\operatorname{Im}\left(  \psi^{\dagger}\left\{  \widehat{A},\widehat
{\mathcal{H}}\right\}  \psi\right)  +\frac{1}{\hbar}\operatorname{Im}\left(
\psi^{\dagger}\left[  \widehat{A},\widehat{\mathcal{H}}\right]  \psi\right)
\label{13}%
\end{equation}
with $\left\{  \widehat{a},\widehat{b}\right\}  =\widehat{a}\,\widehat
{b}+\widehat{b}\,\widehat{a}$, and, by integration over the whole space, we
get%
\begin{equation}
\int\text{d}^{3}r\operatorname{Im}\left(  \psi^{\dagger}\left\{  \widehat
{A},\widehat{\mathcal{H}}\right\}  \psi\right)  =0\text{.}\label{14}%
\end{equation}

The time derivative of $\left(  \widehat{A}\right)  $ is composed of two
parts, concerning two different physical processes: we respectively recognize
in Eq. \ref{13} the divergence of the current and the source term $G$
associated to $\widehat{A}$%
\begin{equation}
\mathbf{\nabla\cdot J}_{A}=-\,\frac{1}{\hbar}\operatorname{Im}\left(
\psi^{\dagger}\left\{  \widehat{A},\widehat{\mathcal{H}}\right\}  \psi\right)
=-\,\frac{1}{\hbar}\operatorname{Im}\left(  \psi^{\dagger}\left\{  \widehat
{A},\widehat{\mathcal{H}}-\mathcal{U}\right\}  \psi\right)  \text{,}\label{15}%
\end{equation}
where any real potential $\mathcal{U}$ vanishes when taking the imaginary part
of the anticommutator, and%
\begin{equation}
G=\frac{1}{\hbar}\operatorname{Im}\left(  \psi^{\dagger}\left[  \widehat
{A},\widehat{\mathcal{H}}\right]  \psi\right)  \text{.}\label{16}%
\end{equation}

The above procedure has two advantages: first, we have expressed in a general
form all the quantities entering Eq. \ref{3} through commutators and
anticommutators; then we have related the probability-current expression
directly to the local properties of its corresponding operator, without taking
into account a closed system (such a procedure does not automatically imply
that the integral of $\mathbf{\nabla\cdot J}_{A}$ over the crystal only is
zero). The choice of considering open systems makes the current operator
involve Dirac distributions to deal properly with possible discontinuities at
the boundaries of a subsystem. It has to be noted that it is always possible
to include the source $G$ term in the form of a current $\mathbf{J}_{G}$,
$G=\mathbf{\nabla\cdot J}_{G}$ so that the conservation equation becomes%
\begin{equation}
\frac{\partial}{\partial t}\left(  \widehat{A}\right)  +\mathbf{\nabla\cdot
}\left(  \mathbf{J}_{A}-\mathbf{J}_{G}\right)  =\frac{\partial}{\partial
t}\left(  \widehat{A}\right)  +\mathbf{\nabla\cdot}\mathcal{J}\mathbf{=}%
0\label{H 10}%
\end{equation}
where $\mathcal{J}=\mathbf{J}_{A}-\mathbf{J}_{G}$ is divergence-free in
steady-state regime. For instance, if we look for $\mathbf{J_{G}=}$
$\mathbf{\nabla}U_{G}$, the potential $U_{G}$ is a solution of the Laplacian
problem $\Delta U_{G}=G$. Moreover, adding to $\mathbf{J}_{G}$ the term
$\mathbf{\nabla\times}\mathcal{A}_{G}$, where $\mathcal{A}_{G}$ is an
arbitrary vector field, does not affect the conservation equation. At this
stage, the boundary conditions are not under control. Shi et al.\cite{Shi06}
observe that it might often happen that
\begin{equation}
\int_{\mathbb{V}}\text{d}^{3}r\;G=0\label{H 20}%
\end{equation}
where the integration is performed over the volume of the system $\left(
\mathbb{V}\right)  $. Then%
\begin{equation}
\int_{\mathbb{V}}\text{d}^{3}r\;G=\int_{\mathbb{V}}\text{d}^{3}%
r\;\mathbf{\nabla\cdot J}_{G}=\int_{\mathbb{S}}\mathbf{J}_{G}\cdot
\text{d}\mathbf{s}=0\label{H 30}%
\end{equation}
where the volume integral is changed into a surface integral through
Ostrogradski's theorem (here $\mathbb{S}$ is the surface limiting $\mathbb{V}
$ and d$\mathbf{s}$ is the surface element along the normal to $\mathbb{S}$).
Such a relation is obviously satisfied provided that $\mathbf{J}_{G}\cdot
$d$\mathbf{s}=0$, i.e., provided that $\mathbf{J}_{G}$ is a tangential vector
to $\mathbb{S}$, which is \textquotedblleft physically\textquotedblright%
\ reasonable. Shi et al. further assume that $\mathbf{J_{G}}$
\textquotedblleft is a material property that should vanish outside the
sample\textquotedblright: this is a more restrictive and questionable
hypothesis. For instance in the case of a magnetic field, the effect of the
associated vector potential cannot \textit{a priori} be overlooked outside the
sample. Anyway, let us assume that $\mathbf{J_{G}=0}$ at the surface
$\mathbb{S}$. Following the calculation by Shi et al., it is straightforward
to show, after partial integration where the boundary contribution cancels,
that%
\begin{equation}
\int\text{d}y\ \text{d}z\ \text{d}x\;x\left(  \frac{\partial J_{G,x}}{\partial
x}+\frac{\partial J_{G,y}}{\partial y}+\frac{\partial J_{G,z}}{\partial
z}\right)  =-\int\text{d}^{3}r\;J_{G,x}\label{H 40}%
\end{equation}
where $J_{G,x}$, $J_{G,y}$, and $J_{G,z}$ are the Cartesian components of
$\mathbf{J_{G}}$. Then%
\begin{align}
\int\text{d}^{3}r\;\mathbf{J}_{G} &  =-\int\text{d}^{3}r\;\mathbf{r}%
\,\mathbf{\nabla\cdot J}_{G}=-\int\text{d}^{3}r\;\mathbf{r}\,G\nonumber\\
&  =-\frac{1}{\hbar}\int\text{d}^{3}r\;\mathbf{r}\,\operatorname{Im}\left(
\psi^{\dagger}\left[  \widehat{A},\widehat{\mathcal{H}}\right]  \psi\right)
=-\frac{1}{\hbar}\int\text{d}^{3}r\;\operatorname{Im}\left(  \psi^{\dagger
}\mathbf{r}\left[  \widehat{A},\widehat{\mathcal{H}}\right]  \psi\right)
\text{.}\label{H 50}%
\end{align}

It is easy to check that, provided that $\left[  \widehat{A},\mathbf{r}%
\right]  =0$,%
\begin{equation}
\mathbf{r}\left[  \widehat{A},\widehat{\mathcal{H}}\right]  =\left[
\widehat{A}\mathbf{r},\widehat{\mathcal{H}}\right]  -i\hbar\widehat{v}%
\widehat{A}\text{,}\label{H 60}%
\end{equation}
where $\left[  \mathbf{r},\widehat{\mathcal{H}}\right]  =i\hbar\widehat{v}$.
Thus
\begin{align}
\int\text{d}^{3}r\;\mathbf{J}_{G} &  =-\frac{1}{\hbar}\int\text{d}%
^{3}r\;\operatorname{Im}\left(  \psi^{\dagger}\left[  \widehat{A}%
\mathbf{r},\widehat{\mathcal{H}}\right]  \psi\right)  +\int\text{d}%
^{3}r\;\operatorname{Re}\left(  \psi^{\dagger}\widehat{v}\widehat{A}%
\psi\right) \nonumber\\
&  =-\frac{1}{\hbar}\int\text{d}^{3}r\;\operatorname{Im}\left(  \psi^{\dagger
}\left[  \widehat{A}\mathbf{r},\widehat{\mathcal{H}}\right]  \psi\right)
+\int\text{d}^{3}r\;\widetilde{\mathbf{J}}_{A}\text{.}\label{H 70}%
\end{align}

Here, $\widetilde{\mathbf{J}}_{A}$ is the canonical current defined as%
\begin{equation}
\widetilde{\mathbf{J}}_{A}=\operatorname{Re}\left(  \psi^{\dagger}%
\,\widehat{v}\,\widehat{A}\,\psi\right)  =\psi^{\dagger}\frac{\widehat
{v}\,\widehat{A}+\widehat{A}\,\widehat{v}}{2}\psi\text{.}\label{H 80}%
\end{equation}

According to Eq. \ref{6}, we can write%
\begin{equation}
\int\text{d}^{3}r\;\mathbf{J}_{G}=-\int\text{d}^{3}r\;\left[  \frac
{\text{d}\left(  \widehat{A}\mathbf{r}\right)  }{\text{d}t}\mathbf{-}%
\widetilde{\mathbf{J}}_{A}\right]  =\int\text{d}^{3}r\;\left[  \widetilde
{\mathbf{J}}_{A}-\frac{\text{d}\left(  \widehat{A}\mathbf{r}\right)
}{\text{d}t}\right]  \text{.}\label{H 90}%
\end{equation}

Shi et al. define the \textit{effective} current density as $\overline
{\mathbf{J}}_{G}$
\[
\overline{\mathbf{J}}_{G}=\widetilde{\mathbf{J}}_{A}-\frac{\text{d}\left(
\widehat{A}\mathbf{r}\right)  }{\text{d}t}\text{.}%
\]
We have the two following relations which define respectively the total
current $\mathcal{J}$ and the\textit{\ effective} total current $\overline
{\mathcal{J}}$
\begin{subequations}
\label{H 100}%
\begin{align}
\mathcal{J}  & \mathcal{=}\mathbf{J}_{A}-\mathbf{J}_{G}\text{,}\label{H 100 a}%
\\
\overline{\mathcal{J}}  & \mathcal{=}\mathbf{J}_{A}-\overline{\mathbf{J}}%
_{G}=\frac{\text{d}\left(  \widehat{A}\mathbf{r}\right)  }{\text{d}t}+\left(
\mathbf{J}_{A}-\widetilde{\mathbf{J}}_{A}\right)  \text{.}\label{H 100 b}%
\end{align}

Provided $\mathbf{J}_{A}-\widetilde{\mathbf{J}}_{A}=0$, i.e. when making the
confusion between the canonical and the true currents (which is justified only
for Hamiltonians up to second order in $\widehat{\mathbf{p}}$, see Sec. III),
the effective total current$\ $becomes $\overline{\mathcal{J}}=$d$\left(
\widehat{A}\mathbf{r}\right)  /$d$t$, which is Eq. 5 in the papers by Shi et
al.\cite{Shi06} and also by Zhang et al.,\cite{Zhang08} and is the cornerstone
of their further calculations. After a careful analysis, this relation appears
to be derived under very special conditions so that it cannot be general.
Moreover, the meaning of the so-called effective currents and their
relationship with the true currents are not clear. Their use to tackle local
transport equations is not justified.

\section{Probability current of an effective Hamiltonian}

\subsection{Formulation of the general $n^{th}$-order Hamiltonian}

Considering \textit{effective }Hamiltonians, we deal with general expressions
given by momentum series expansions, i.e., constructed from the energy
expressed as wave-vector-component series expansion after the substitution
$\left\{  k\longrightarrow-i\mathbf{\nabla}\right\}  $. We write the effective
Hamiltonian $\widehat{H}_{eff}$ as follows:
\end{subequations}
\begin{equation}
\widehat{H}_{eff}=\widehat{H}_{\mathbf{p}}+V\left(  \mathbf{r}\right)
\label{a 37}%
\end{equation}
where $V\left(  \mathbf{r}\right)  $ is a potential which may be the potential
of a single barrier or the one of a superlattice, for example, $\widehat
{H}_{\mathbf{p}}$ is such that%
\begin{equation}
\widehat{H}_{\mathbf{p}}=\sum_{n}\underset{\overset{l(k)\in\left\{
x,y,z\right\}  }{k=1,...,n}}{\sum}c_{l(1),l\left(  2\right)  ,}..._{,l(n)}%
\widehat{p}_{l(1)}...\widehat{p}_{l(n)}=\sum_{n}\widehat{H}^{\left(  n\right)
}\label{37}%
\end{equation}
where $\widehat{p}_{l(k)}$ is the momentum operator associated to the $l(k)$
Cartesian coordinate and where $c_{l(1)},...,_{l(n)}$ are Hermitian matrices
which are invariant under permutation of the subscripts. The abstract form of
Eq. \ref{37} allows us to perform easy calculations. In Sec. \ref{AppSOI} we
show how to handle such a general expression to deal with concrete situations.

Formally, we perform the identification%
\begin{equation}
\underset{\alpha}{\underbrace{c_{x}...c_{x}}}\underset{\beta}{\,\underbrace
{c_{y}...c_{y}}\,}\underset{\gamma}{\underbrace{c_{z}...c_{z}}}=\underset
{\alpha}{c_{\underbrace{x...x}}}\underset{\beta}{_{,\,\underbrace{y...y}}%
}\underset{\gamma}{_{,\,\underbrace{z...z}}}\label{58}%
\end{equation}
where $\alpha$, $\beta$, and $\gamma$ are integers. We obtain%
\begin{equation}
\widehat{H}^{(n)}=\left(  c_{x}\widehat{p}_{x}+c_{y}\widehat{p}_{y}%
+c_{z}\widehat{p}_{z}\right)  ^{n}\text{.}\label{52}%
\end{equation}

Given Eqs. \ref{37},\ \ref{58}, and \ref{52}, let us note that only terms such
as $c_{xx}$ or $c_{xy}$ (for $n=2$) are meaningful, a term such as $c_{x}$
being only a trick in the calculation.

Alternatively, one can write%
\begin{equation}
\widehat{H}^{(n)}=\underset{\alpha+\beta+\gamma=n}{\sum}c^{\alpha\beta\gamma
}\,\,\widehat{p}_{x}^{\,\alpha}\,\,\widehat{p}_{y}^{\,\beta}\,\,\widehat
{p}_{z}^{\,\gamma}\label{55}%
\end{equation}
with%
\begin{equation}
c^{\alpha\beta\gamma}=\frac{n!}{\alpha!\beta!\gamma!}c_{x}^{\alpha}%
c_{y}^{\beta}c_{z}^{\gamma}\label{57}%
\end{equation}

We are now in a position to tackle the problem of velocity, first when the
Hamiltonian $\widehat{H}$ takes into account the SOI, and, second, when the
Hamiltonian $\widehat{H}_{eff}$ is an effective Hamiltonian.

\subsection{Velocity operator in presence of SOI interaction\label{AppSOI}}

It is usually admitted that the velocity operator $\widehat{\mathbf{v}}$ is
equal to $\partial\widehat{\mathcal{H}}/\partial\widehat{\mathbf{p}}$ whatever
the Hamiltonian $\widehat{\mathcal{H}}$.\ However, to the best of our
knowledge, the derivation can be found only when $\widehat{\mathcal{H}}$ is
quadratic in $\widehat{\mathbf{p}}$. Therefore a general derivation, in
particular in the case of effective Hamiltonians, is mandatory. We start from
Ehrenfest's theorem (valid whatever the Hamiltonian $\widehat{\mathcal{H}}$)%
\begin{equation}
\langle\widehat{\mathbf{v}}\rangle=\frac{\text{d}\langle\widehat{\mathbf{r}%
}\rangle}{\text{d}t}=\frac{i}{\hbar}\left\langle \left[  \widehat{\mathcal{H}%
},\ \widehat{\mathbf{r}}\right]  \right\rangle \label{155}%
\end{equation}

If $\left(  i/\hbar\right)  \left\langle \left[  \widehat{\mathcal{H}%
},\ \widehat{\mathbf{r}}\right]  \right\rangle =\left\langle \partial
\widehat{\mathcal{H}}/\partial\mathbf{p}\right\rangle $, then $\widehat
{\mathbf{v}}=\partial\widehat{\mathcal{H}}/\partial\widehat{\mathbf{p}}$
because two linear operators which have the same mean values over all possible
states are equal: $\left\langle \widehat{A}\right\rangle =\left\langle
\widehat{B}\right\rangle $ implies that $\widehat{A}=\widehat{B}%
$.\cite{Messiah540} Practically, it is enough to show that $\left(
i/\hbar\right)  \left[  \widehat{\mathcal{H}},\ \widehat{\mathbf{r}}\right]
=\partial\widehat{\mathcal{H}}/\partial\widehat{\mathbf{p}}$ to prove that
$\widehat{\mathbf{v}}=\partial\widehat{\mathcal{H}}/\partial\widehat
{\mathbf{p}}$.

First, considering the case $\widehat{\mathcal{H}}=\widehat{H}_{0}$ which
contains a $\widehat{H}^{\left(  2\right)  }$ term (Eq. \ref{H0}), $\left(
i/\hbar\right)  \left[  \widehat{H}_{0},\ \widehat{\mathbf{r}}\right]
=\left(  \hbar/im\right)  \widehat{\mathbf{p}}\mathbf{=\partial}\widehat
{H}_{0}/\partial\widehat{\mathbf{p}}$, for a system described by an
Hamiltonian\ quadratic versus momentum components, and we obtain the velocity
$\widehat{\mathbf{v}}_{0}$:%
\begin{equation}
\widehat{\mathbf{v}}_{0}=\frac{\partial\widehat{H}_{0}}{\partial
\widehat{\mathbf{p}}}\text{.}\label{28bis}%
\end{equation}

Second, we have to check that this relation still holds in the presence of SOI
where the Hamiltonian is $\widehat{\mathcal{H}}=\widehat{H}=\widehat{H}%
_{0}+\widehat{H}_{SO}$ (Eq. \ref{H}). In other words we want to show that%
\begin{equation}
\widehat{\mathbf{v}}=\frac{\partial\widehat{H}}{\partial\widehat{\mathbf{p}}%
}\text{.}\label{a 100}%
\end{equation}

We know that $\widehat{\mathbf{v}}_{0}=\left(  i/\hbar\right)  \left[
\widehat{H}_{0},\ \widehat{\mathbf{r}}\right]  =\mathbf{\partial}\widehat
{H}_{0}/\partial\widehat{\mathbf{p}}$. To show that Eq. \ref{a 100} is valid,
it is enough to show that $\widehat{\mathbf{v}}_{SO}=\partial\widehat{H}%
_{SO}/\partial\widehat{\mathbf{p}}$, which will give $\widehat{\mathbf{v}%
}\mathbf{=\partial}\widehat{H}\mathbf{/\partial\widehat{\mathbf{p}}}$ with
$\widehat{\mathbf{v}}\mathbf{=\widehat{\mathbf{v}}}_{0}+\widehat{\mathbf{v}%
}_{SO}$. A straightforward calculation yields%
\begin{equation}
\widehat{\mathbf{v}}_{SO}=\frac{i}{\hbar}[\widehat{H}_{SO},\ \widehat
{\mathbf{r}}]=\dfrac{\mathbf{\hbar}}{4m^{2}c^{2}}\left(  \widehat
{\mathbf{\sigma}}\mathbf{\times\nabla}\mathfrak{U}\right)  =\frac
{\partial\widehat{H}_{SO}}{\partial\widehat{\mathbf{p}}}\label{30}%
\end{equation}
which proves Eq.~\ref{a 100}:~the derivative of the Hamiltonian, with respect
to the momentum operator, still provides a suitable definition of the velocity
when the SOI term is taken into account.

\subsection{Velocity operator with an effective Hamiltonian $\widehat{H}%
_{eff}$}

We generalize the results obtained in Sec. \ref{AppSOI}, to the case of a
generic effective Hamiltonian $\widehat{\mathcal{H}}=\widehat{H}_{eff}$. Again
we exploit Ehrenfest's theorem, as written in Eq. \ref{155}. Considering for
instance the $x$ component, we verify that%
\begin{equation}
\left[  \widehat{H}^{\left(  n\right)  },\ x\right]  =\underset{\alpha
+\beta+\gamma=n}{\sum}c^{\alpha\beta\gamma}\left(  \frac{\hbar}{i}\right)
\alpha\,\widehat{p}_{x}^{\alpha-1}\,\widehat{p}_{y}^{\,\beta}\,\widehat{p}%
_{z}^{\,\gamma}=\frac{\hbar}{i}\frac{\partial\widehat{H}^{\left(  n\right)  }%
}{\partial\widehat{p}_{x}}\label{174}%
\end{equation}
or%
\begin{equation}
\frac{i}{\hbar}\left[  \widehat{H}_{\mathbf{p}},\ \widehat{\mathbf{r}}\right]
=\frac{i}{\hbar}\left[  \widehat{H}_{\mathbf{eff}},\ \widehat{\mathbf{r}%
}\right]  =\frac{\partial\widehat{H}_{eff}}{\partial\widehat{\mathbf{p}}%
}\label{a 174}%
\end{equation}
which proves that%
\begin{equation}
\widehat{\mathbf{v}}=\frac{\partial\widehat{H}_{eff}}{\partial\widehat
{\mathbf{p}}}\text{.}\label{b 174}%
\end{equation}

Using\ Eqs. \ref{58}-\ref{57}, it is then easy to calculate the $j$ component
$\widehat{v}_{j}^{(n)}$ $\left(  j=x,\,y,\,z\right)  $ of the velocity
operator $\widehat{\mathbf{v}}^{\left(  n\right)  }$ associated to
$\widehat{H}^{(n)}$:
\begin{equation}
\widehat{v}_{j}^{\left(  n\right)  }=\frac{\partial\widehat{H}^{(n)}}%
{\partial\widehat{p}_{j}}=nc_{j}\left(  c_{x}\widehat{p}_{x}+c_{y}\widehat
{p}_{y}+c_{z}\widehat{p}_{z}\right)  ^{n-1}\text{.}\label{53}%
\end{equation}

We introduce the scalar product between the momentum $\widehat{\mathbf{p}}$
and the velocity operator $\widehat{\mathbf{v}}^{\left(  n\right)  }$%
\begin{equation}
\widehat{p}_{x}\widehat{v}_{x}^{(n)}+\widehat{p}_{y}\widehat{v}_{y}%
^{(n)}+\widehat{p}_{z}\widehat{v}_{z}^{(n)}=n\left(  c_{x}\widehat{p}%
_{x}+c_{y}\widehat{p}_{y}+c_{z}\widehat{p}_{z}\right)  ^{n}=n\widehat{H}%
^{(n)}\text{.}\label{a 140}%
\end{equation}

With this notation, $\widehat{\mathbf{v}}_{0}$, introduced in the paragraph
\ref{AppSOI}, is such that $\widehat{\mathbf{v}}_{0}=\widehat{\mathbf{v}%
}^{\left(  2\right)  }$. Eq. \ref{a 140} means that%
\begin{equation}
\widehat{\mathbf{p}}\cdot\widehat{\mathbf{v}}^{\left(  n\right)  }%
=n\widehat{H}^{(n)}\label{a 160}%
\end{equation}
and eventually
\begin{equation}
\widehat{H}_{eff}\psi=\left(  \widehat{\mathbf{p}}\cdot\sum\limits_{n}\frac
{1}{n}\widehat{\mathbf{v}}^{(n)}\right)  \psi+V\psi=E\psi\text{.}\label{62}%
\end{equation}

As pointed out in Sec. \ref{sec Gen}, we are allowed to define current
operators in open systems provided that we properly take into account their
boundary conditions. In Appendix \ref{AppCurr}, we show that performing the
proper symmetrization according to the following rule (Eq. \ref{860}), we find
a probability current operator that for the $j^{th}$-Cartesian component
reads:%
\begin{equation}
\widehat{J}_{j}\left(  \mathbf{r}_{0}\right)  =\sum_{n}\widehat{J}%
_{j}^{\,\left(  n\right)  }\left(  \mathbf{r}_{0}\right) \label{860 bis}%
\end{equation}%
\begin{equation}
\widehat{J}_{j}^{\,\left(  n\right)  }\left(  \mathbf{r_{0}}\right)
=\underset{\overset{l(k)\in\left\{  x,y,z\right\}  }{k=1,..,n-1}}{\sum
}c_{j,l(1),}..._{,l(n-1)}\left[  \delta_{\mathbf{r}_{0}}\widehat{p}%
_{l(1)}...\widehat{p}_{l(n-1)}+\widehat{p}_{l(1)}\delta_{\mathbf{r}_{0}%
}...\widehat{p}_{l(n-1)}+....+\widehat{p}_{l(1)}...\widehat{p}_{l(n-1)}%
\delta_{\mathbf{r}_{0}}\right] \label{860}%
\end{equation}
where $\delta_{\mathbf{r}_{0}}=\delta(\mathbf{r}-\mathbf{r}_{0})$ is the Dirac
distribution. We must still verify that the divergence of the current,
calculated with the operator defined by Eq. \ref{860}, satisfies the
conservation equation for the density of probability (Eq. \ref{15} when
$\widehat{A}$ is the identity). It is straightforward to show (see Appendix
\ref{AppCurr}) that the divergence of the probability current can be written
as:%
\begin{equation}
\mathbf{\nabla}\cdot\mathbf{J}\left[  \psi\right]  =\sum_{n}\mathbf{\nabla
}\cdot\mathbf{J}^{\left(  n\right)  }\left[  \psi\right]  =-\frac{2}{\hbar
}\;\operatorname{Im}\sum_{n}\;\sum_{j=\left\{  x,y,z\right\}  }\;\;\sum
_{\overset{l(k)\in\left\{  x,y,z\right\}  }{k=1,..,n-1}}c_{j,l(1),}%
..._{,l(n-1)}\left(  \psi\left\vert \widehat{p}_{j}\widehat{p}_{l(1)}%
...\widehat{p}_{l(n-1)}\right\vert \psi\right) \label{87}%
\end{equation}
where we again use the notations $\left(  \psi\left\vert \widehat{A}%
\psi\right.  \right)  =\psi^{\dagger}\widehat{A}\psi$ as in Ref.
\onlinecite{Hoai09}%
. Then, we recover all the terms of Eq. \ref{15}, so that Eq. \ref{860} indeed
provides a correct definition of the current operator. Obviously, adding a
term proportional to the curl of any vector field would not affect the result.
Such a definition of $\mathbf{\hat{J}}$ provides an unambiguous and general
tool for evaluating the probability current. Provided the Hamiltonian of the
whole system is known, this probability-current operator guarantees the
requirements of the continuity equation.

Now it is useful to introduce the Hermitian symmetrized velocity operator%
\begin{equation}
\widehat{\mathfrak{v}}_{j}^{\,\left(  n\right)  }(\mathbf{r_{0}}%
)=\underset{\overset{l(k)\in\left\{  x,y,z\right\}  }{k=1,..,n-1}}{\frac{n}%
{2}\sum}c_{j,l(1),}..._{,l(n-1)}\left[  \delta_{\mathbf{r}_{0}}\widehat
{p}_{l(1)}...\widehat{p}_{l(n-1)}+\widehat{p}_{l(1)}...\widehat{p}%
_{l(n-1)}\delta_{\mathbf{r}_{0}}\right] \label{8600}%
\end{equation}

For example for $n\geq2$, the comparison between Eqs. \ref{860} and \ref{8600}
clearly shows that $\widehat{J}_{j}^{\,\left(  n\right)  }(\mathbf{r_{0}})$
contains $n-2$ extra terms, which are straightforwardly obtained from
$\partial\widehat{H}_{eff}/\partial\widehat{\mathbf{p}}$. For instance, with
$\widehat{H}_{eff}\equiv\widehat{p}^{n}$, we have $\partial\widehat{H}%
_{eff}/\partial\widehat{\mathbf{p}}\equiv n\widehat{p}^{n-1}$, so that
$\widehat{\mathfrak{v}}^{\left(  n\right)  }(\mathbf{r_{0}})\equiv\left(
n/2\right)  \left(  \delta_{\mathbf{r}_{0}}\widehat{p}^{\,n-1}+\widehat
{p}^{\,n-1}\delta_{\mathbf{r}_{0}}\right)  $, whereas $\widehat{J}^{\,\left(
n\right)  }\left(  \mathbf{r_{0}}\right)  \equiv\left(  \delta_{\mathbf{r}%
_{0}}\widehat{p}^{n-1}+\widehat{p}\delta_{\mathbf{r}_{0}}\widehat{p}%
^{n-2}+...+\widehat{p}^{n-1}\delta_{\mathbf{r}_{0}}\right)  $. As shown in
Ref.
\onlinecite{Hoai09}%
, extra terms are specially important for evanescent waves.\ Therefore, in the
following we deal with tunneling problems.

\section{B%
\lowercase{en}%
D%
\lowercase{aniel}%
-D%
\lowercase{uke}%
-like formulation and boundary conditions\label{BDD}}

We stress that the central question when defining the current operators and
related quantities is the proper definition of the system and of its
boundaries. Dealing with heterojunctions, where each bulk medium is described
by the relevant Hamiltonian, requires defining proper matching conditions at
the boundaries. In this sense, the BDD Hamiltonian\cite{BenDaniel} is the
simplest smart approach that allows solution of the Schr\"{o}dinger equation
over the whole space while it guaranties the conservation of the probability
current at the interface. The principle is the following. Let us consider a
one-dimensional problem and two different media for $x<0$ and $x>0$. Each
medium is characterized by its own Hamiltonian. The question is to find a
solution of the Schr\"{o}dinger equation, made of eigenfunctions of the
relevant band of the two bulk materials, which ensures the continuity of the
probability current at the origin. In this sense, the problem is analogous to
a scattering problem, where the wave functions are determined only at some
distance of the scattering potential. Proper matching conditions relevant to
the extension of the bulk envelope functions at the origin will allow one to
determine the envelope function over the whole space. For that, BBD propose
writing an Hamiltonian over the whole space as $\widehat{p}_{x}\left[
1/2m\left(  x\right)  \right]  \widehat{p}_{x}+V\left(  x\right)  $ where
$m\left(  x\right)  \ $is the effective mass in each medium. The integration
of this BDD Hamiltonian around the boundary automatically ensures the
continuity of the probability current of Eq. \ref{1}, provided that
$\psi\left(  x\right)  $ and $\left[  1/m\left(  x\right)  \right]  \left[
\partial\psi/\partial x\right]  $ are continuous.

Now, consider two regions $\left(  1\right)  $ and $\left(  2\right)  $ and
assume that each region is made of a given crystalline material. We look for
the envelope function, solution of the Schr\"{o}dinger equation, which is made
from plane waves which are eigenstates of the crystal, inside each material.
Observe that, near the interface, the crystal periodicity is broken so that
the true Hamiltonian and the true eigenfunctions will become involved. The
principle is then to define proper matching conditions applying to the
prolongation\ of the envelope function at the origin. For that purpose, we
consider a volume $\mathbb{V}$, limited by a surface $\mathbb{S}$, that
surrounds an interface portion. Similarly to the BDD technique, we start from
Eq. \ref{62} and we integrate the Schr\"{o}dinger equation over $\mathbb{V}$.
Using Ostrogradski's theorem, when $\mathbb{V}$ tends to zero, we obtain%
\begin{equation}
\underset{\mathbb{V}\longrightarrow0}{\lim}\int_{\mathbb{S}}\left(  \sum
_{n}\frac{1}{n}\mathbf{v}^{(n)}\psi\right)  \cdot\text{d}\mathbf{s}%
=0\label{64}%
\end{equation}
where d$\mathbf{s}$ is normal to the surface $\mathbb{S}$.

For a one dimensional case with the interface at the origin, Eq. \ref{64}
becomes:%
\begin{equation}
\underset{\varepsilon\rightarrow0}{\lim}\left[  \sum_{n}\frac{1}{n}v^{(n)}%
\psi\right]  _{-\varepsilon}^{+\varepsilon}=0\label{65}%
\end{equation}

Let us again emphasize that no information is obtained on the true wave
function near the origin. Eq. \ref{65} does not ensure either the continuity
of the envelope function or the existence of derivatives \textit{at} the interface.

As an illustration, let us consider the case of a Rashba Hamiltonian%
\begin{equation}
\hat{H}_{eff}=a\widehat{p}+b\widehat{p}^{2}\label{150}%
\end{equation}
where $a$ and $b$ are two Hermitian matrices. According to Eq. \ref{65}, we
can write down the first continuity condition as follows:%
\begin{equation}
\left[  a\psi+b\widehat{p}\psi\right]  _{-\varepsilon}^{+\varepsilon
}=0\text{.}\label{152}%
\end{equation}

Using this condition to solve the problem, and adding \textit{a priori} the
continuity of the envelope function at the interface as a second condition, we
verify that the probability current is indeed continuous at the interface:%
\begin{equation}
J\left[  \psi\right]  =\langle\psi\mid a+b\widehat{p}\mid\psi\rangle
+c.c.\label{154}%
\end{equation}

Then, the jump of the derivative of the wave function at the interface is
determined by%
\begin{equation}
\left[  b\widehat{p}\psi\right]  _{-\varepsilon}^{+\varepsilon}=-\left[
a\right]  _{-\varepsilon}^{+\varepsilon}\psi(0)\text{.}\label{153}%
\end{equation}
It is clear then that the BDD approach, introduced to solve a problem with a
quadratic Hamiltonian, is also suitable to obtain a solution when a Rashba
contribution is added; then we can say that up to the second order in the
momentum-power series expansion of the Hamiltonian, the continuity of a
\textquotedblleft generalized velocity\textquotedblright\ (see Eq. \ref{65})
and the continuity of the wave function at the interface imply the
conservation of the probability current at this point. Remarkably, the
boundary conditions that we need to solve the problem drastically change when
moving to the case of a DP Hamiltonian with cubic terms. The crucial point,
that we address in the following, is that we cannot make any hypothesis about
the continuity of the wave function because, if we need to ensure
probability-current conservation at an interface, we must accept an envelope
function $\psi$ which is no longer continuous.

To give an insight into the expression of the current operator and into the
conservation of the probability current, let us again come back to an
interface between two semi-infinite one-dimensional media $\left(  1\right)  $
and $\left(  2\right)  $. In each bulk crystal, the relevant Hamiltonian is%
\begin{equation}
\widehat{H}_{r}=\sum\limits_{n}\widehat{H}_{r}^{\left(  n\right)  }%
+V_{r}\label{H_j}%
\end{equation}
with%
\begin{equation}
\widehat{H}_{r}^{\left(  n\right)  }=\gamma_{r}^{\left(  n\right)  }%
\widehat{p}^{n}\label{101}%
\end{equation}
with $r=1$ or $r=2$ depending on wether $x<0$ or $x>0$. $\widehat{H}_{r}$
admits the eigenfunctions $\varphi_{r}$, associated to the fixed energy $E$
which verify%
\begin{equation}
\widehat{H}_{r}\varphi_{r}=E\varphi_{r}\text{.}\label{102}%
\end{equation}

Near the heterojunction, the spatial periodicity is broken, so that over a few
Wigner-Seitz cells, the electron states are no longer pure Bloch states. We
consider two coordinates, $-w_{1}$ and $w_{2}$, so that, in the regions
$\left]  -\infty,\,-w_{1}\right]  $ and $\left[  w_{2},\,+\infty\right[  $ the
electronic structure remains unaffected. In these regions, the Hamiltonian
writes%
\begin{equation}
\widehat{\mathfrak{H}}=\Theta\left(  -x-w_{1}\right)  \widehat{H}_{1}%
+\Theta\left(  x-w_{2}\right)  \widehat{H}_{2}\label{103}%
\end{equation}
where $\Theta\left(  x\right)  $ is the Heaviside function. In the vicinity of
the heterojunction, $\left]  -w_{1},w_{2}\right[  $, the form of the
Hamiltonian and of the wave functions are not explicitly known. We consider a
wave function $\psi$ which is an eigenstate of the Hamiltonian over the whole
space at energy $E$. We expect that, over the domain $\left]  -\infty
,-w_{1}\right]  $ $\cup$ $\left[  w_{2},+\infty\right[  $, $\psi$ coincides
with $\Psi=\Theta\left(  -x-w_{1}\right)  \varphi_{1}+\Theta\left(
x+w_{2}\right)  \varphi_{2}$. We thus expect the following relation to be
satisfied%
\begin{equation}
\left\langle \varphi_{1}\Theta\left(  -x\right)  +\varphi_{2}\Theta\left(
x\right)  \left\vert \widehat{\mathfrak{H}}\right\vert \Psi\right\rangle
=E\left[  \left\langle \varphi_{1}\mid\varphi_{1}\right\rangle _{\left(
1\right)  }+\left\langle \varphi_{2}\mid\varphi_{2}\right\rangle _{\left(
2\right)  }\right] \label{H_goth}%
\end{equation}
where $\left\langle \mid\right\rangle _{\left(  r\right)  }$ means summation
over the bulk part of region $\left(  r\right)  $. Observe that:%
\begin{align}
\widehat{H}_{1}^{\left(  n\right)  }\Psi &  =\gamma_{1}^{\left(  n\right)
}\widehat{p}^{n}\Theta\left(  -x-w_{1}\right)  \varphi_{1}+V_{1}\Theta\left(
-x-w_{1}\right)  \varphi_{1}\nonumber\\
&  =\Theta\left(  -x-w_{1}\right)  \left(  \gamma_{1}^{\left(  n\right)
}\widehat{p}^{n}+V_{1}\right)  \varphi_{1}\nonumber\\
&  +i\hbar\gamma_{1}^{\left(  n\right)  }\left[  \delta\left(  x+w_{1}\right)
\widehat{p}^{n-1}+\widehat{p}\delta\left(  x+w_{1}\right)  \widehat{p}%
^{n-2}+...+\widehat{p}^{n-1}\delta\left(  x+w_{1}\right)  \right]  \varphi
_{1}\nonumber\\
&  =\Theta\left(  -x-w_{1}\right)  \left(  \gamma_{1}^{\left(  n\right)
}\widehat{p}^{n}+V_{1}\right)  \varphi_{1}+i\hbar\widehat{J}_{1}^{\left(
n\right)  }\varphi_{1}=\Theta\left(  -x-w_{1}\right)  E\varphi_{1}%
+i\hbar\widehat{J}_{1}^{\left(  n\right)  }\varphi_{1}\label{105}%
\end{align}
and similarly%
\begin{equation}
\widehat{H}_{2}^{\left(  n\right)  }\Psi=\Theta\left(  x+w_{2}\right)
E\varphi_{2}-i\hbar\widehat{J}_{2}^{\left(  n\right)  }\varphi_{2}%
\text{.}\label{106}%
\end{equation}

So that%
\begin{align}
\left\langle \varphi_{1}\Theta\left(  -x\right)  +\varphi_{2}\Theta\left(
x\right)  \left\vert \widehat{\mathfrak{H}}\right\vert \Psi\right\rangle  &
=E\left[  \left\langle \varphi_{1}\mid\varphi_{1}\right\rangle _{\left(
1\right)  }+\left\langle \varphi_{2}\mid\varphi_{2}\right\rangle _{\left(
2\right)  }\right] \nonumber\\
&  +i\hbar\left(  \widehat{J}_{2}^{\left(  n\right)  }\left[  \varphi
_{2}\left(  w_{2}\right)  \right]  -\widehat{J}_{1}^{\left(  n\right)
}\left[  \varphi_{1}\left(  -w_{1}\right)  \right]  \right) \nonumber\\
&  =E\left[  \left\langle \varphi_{1}\mid\varphi_{1}\right\rangle _{\left(
1\right)  }+\left\langle \varphi_{2}\mid\varphi_{2}\right\rangle _{\left(
2\right)  }\right]  \text{.}\label{107}%
\end{align}

This implies that
\begin{equation}
\widehat{J}_{2}^{\left(  n\right)  }\left[  \varphi_{2}\left(  w_{2}\right)
\right]  =\widehat{J}_{1}^{\left(  n\right)  }\left[  \varphi_{1}\left(
-w_{1}\right)  \right]  \text{.}\label{108}%
\end{equation}

The important point is not the conservation relation, which might appear as
physically obvious, but that, in Eqs. \ref{105} and \ref{106}, the symmetrized
current operator is automatically generated in the form derived in Eq.
\ref{860}, providing a physical insight into this mathematical expression.
Because we only deal with wave functions at some distance from the
heterojunction, the continuity of the true wave function at $x=0$ does not
implies the continuity of the envelope function $\Psi$ which may be
discontinuous. This is in line with the considerations of
Harrison.\cite{Harrison} Eq. \ref{65} and \ref{108} generate a set of boundary
conditions relevant to the tunneling problem.

\section{The [110]-oriented G%
\lowercase{a}%
A%
\lowercase{s}
barrier\label{sec GaAs}}

We analyze the case of electron tunneling under normal incidence through a
GaAs $[110]$-oriented barrier, which was shown to be non trivial and solved in
special cases in Ref.
\onlinecite{Hoai09}%
. Hereafter, we apply the tools and boundary conditions presented in this
paper to solve it in a more simple and general manner. We confirm and
generalize the results derived in Ref.
\onlinecite{Hoai09}%
. In particular, we are able to solve the problem of an heterojunction between
a free-electron media and a semiconductor without inversion center, where the
DP field is a step function, which remained puzzling. In the $\left[
110\right]  $ direction, the DP Hamiltonian is:%
\begin{equation}
\widehat{H}_{DP}=\frac{\gamma_{c}}{\hbar^{2}}\widehat{p}^{2}\pm\frac{\gamma
}{2\hbar^{3}}\widehat{p}^{3}\label{AsGa1}%
\end{equation}
where $+$ $\left(  -\right)  $ refers to the up (down)- spin channel quantized
along the DP field direction. We consider as solution a general wavefunction
written as follows:%
\begin{equation}
\psi=\alpha\left(  \psi_{0}+\frac{i\beta}{\hbar\gamma_{c}k^{2}}\gamma
_{c}\widehat{p}\psi_{0}\right)  e^{i\chi z}\label{AsGa2}%
\end{equation}
where $\psi_{0}$ is the zeroth order function that is a solution of the
tunneling problem with energy $E$ and with the potential $V$ when SOI is
turned off. Here $\gamma_{c}k^{2}=E-V$, $\alpha$ and $\beta$ are complex
parameters to be determined, and $\chi$ is a real (see below) wavevector
component which is added to $k$ when SOI is turned on. We have the relations%
\begin{equation}
\widehat{p}\psi=\alpha\left(  \widehat{p}\psi_{0}+i\frac{\beta}{\hbar
\gamma_{c}k^{2}}\gamma_{c}\,\widehat{p}^{2}\,\psi_{0}\right)  e^{i\chi
z}+\hbar\chi\psi\text{,}\label{AsGa3}%
\end{equation}%
\begin{equation}
\widehat{p}^{2}\psi=\hbar^{2}\left(  \frac{E-V}{\gamma_{c}}-\chi^{2}\right)
\psi+2\hbar\chi\widehat{p}\psi\text{.}\label{AsGa4}%
\end{equation}
We calculate the velocity operators from Eq. \ref{53}
\begin{equation}
\frac{1}{2}v^{(2)}+\frac{1}{3}v^{(3)}=\frac{\gamma_{c}}{\hbar^{2}}\widehat
{p}\pm\frac{\gamma}{2\hbar^{3}}\widehat{p}^{2}\label{AsGa5}%
\end{equation}
and, according to Eq. \ref{65}, we find the matching condition%
\begin{equation}
\left[  \gamma_{c}\left(  1\pm\frac{\gamma}{\gamma_{c}}\chi\right)
\widehat{p}\psi\right]  _{-\varepsilon}^{+\varepsilon}=\mp\left[  \frac{1}%
{2}\gamma\hbar\left(  \frac{E-V}{\gamma_{c}}-\chi^{2}\right)  \psi\right]
_{-\varepsilon}^{+\varepsilon}\label{AsGa6}%
\end{equation}
which is a generalization of Eq. (3.50) of Ref.
\onlinecite{Hoai09}%
.

Now, we have to satisfy the conservation of the probability current%
\begin{equation}
J[\psi]=\frac{\gamma_{c}}{\hbar^{2}}\left(  1\pm\frac{\gamma}{\gamma_{c}}%
\chi\right)  \left[  \psi^{\ast}\widehat{p}\psi+\psi\left(  \widehat{p}%
\psi\right)  ^{\ast}\right]  \pm\frac{\gamma}{\hbar}\left(  \frac{E-V}%
{\gamma_{c}}-\chi^{2}\right)  \left\vert \psi\right\vert ^{2}\pm\frac{\gamma
}{2\hbar^{3}}\left\vert \widehat{p}\psi\right\vert ^{2}\text{.}\label{AsGa7}%
\end{equation}

We obtain here an important result: \textit{The envelope function cannot be
continuous at the interface.} Indeed, assume $\psi$ to be continuous. Then,
after Eq. \ref{AsGa6}, we see that the last term in Eq. \ref{AsGa7}, that we
rewrite as $\pm\left(  1/2\hbar^{3}\right)  \left(  \gamma/\gamma_{c}%
^{2}\right)  \gamma_{c}^{2}\left\vert \widehat{p}\psi\right\vert ^{2}$ must be
continuous. This is not possible since $\gamma_{c}\widehat{p}\psi_{0}$ is
continuous (unless $\gamma/\gamma_{c}^{2}$ is almost continuous, which would
be fortuitous).

We have to determine $\psi$ complying the boundary conditions, which is not
simple because the expression providing the current is not a linear function
of $\psi$. However, if we consider $\gamma$ as a first-order quantity and look
for a solution to first order only, the result is surprisingly simple,
as\ shown\ below. From the Schr\"{o}dinger equation - Eq. \ref{AsGa1} -, we
find that $\chi$ verifies%
\begin{equation}
\gamma_{c}\left(  2k\chi+\chi^{2}\right)  \pm\frac{\gamma}{2}\left(
k^{3}+3k^{2}\chi+3k\chi^{2}+\chi^{3}\right)  =0\label{AsGa8}%
\end{equation}
then%
\begin{equation}
\chi\simeq\mp\frac{1}{4}\frac{\gamma}{\gamma_{c}}k^{2}=\mp\frac{1}{4}%
\frac{\gamma}{\gamma_{c}}\frac{E-V}{\gamma_{c}}\text{.}\label{AsGa9}%
\end{equation}

As stated above, $\chi$ is a \textit{real} quantity. For each spin, there are
two others roots of the cubic equation (Eq. \ref{AsGa8}) which are much larger
than the width of the Brillouin zone ; These two roots are of the order of
$\gamma_{c}/\gamma$ which is about $2$ \AA \ (two times the Brillouin zone
width) in GaAs (see Fig. 4 of Ref.
\onlinecite{Hoai09}%
) and have no physical meaning. Note that, the cubic DP term, obtained from
perturbation expansion, only holds for small wave vectors, a few percent of
the Brillouin zone, so that taking into account these two other roots would be
meaningless. From Eq. \ref{AsGa2}, we see that, upon tunneling, the up- and
down- spin electrons undergo opposite phase shifts, which is equivalent to a
precession around the DP-field direction. This would be quite intuitive if the
field were not a complex quantity, and constitutes a prediction which can be
experimentally tested. Let us calculate the current at the interface
$J[\psi\left(  0\right)  ]$ to first order%
\begin{multline}
J[\psi\left(  0\right)  ]=\left\vert \alpha\right\vert ^{2}\frac{\gamma_{c}%
}{\hbar^{2}}\left[  \psi_{0}^{\ast}\,\widehat{p}\,\psi_{0}+\psi_{0}\,\left(
\widehat{p}\,\psi_{0}\right)  ^{\ast}\right]  +\frac{\gamma_{c}}{2\hbar
}\left[  2\left\vert \alpha\right\vert ^{2}\left(  \chi-\operatorname{Im}%
\beta\right)  \pm\frac{\gamma}{\gamma_{c}}\frac{E-V}{\gamma_{c}}\right]
\left\vert \psi_{0}\right\vert ^{2}\\
\left.  +\frac{\gamma_{c}}{\hbar^{3}}\left[  -\left\vert \alpha\right\vert
^{2}\frac{2\gamma_{c}}{E-V}\operatorname{Im}\beta\mp\frac{\gamma}{2\gamma_{c}%
}\right]  \left\vert \widehat{p}\psi_{0}\right\vert ^{2}\right\}
\label{AsGa10}%
\end{multline}
where the values of $\psi_{0}$ and of its derivative are calculated at the
origin. Observe that with the choice%
\begin{equation}
\left\vert \alpha\right\vert ^{2}=1\quad\text{and}\quad\operatorname{Im}%
\beta=-\chi\label{AsGa11}%
\end{equation}
the second and the last terms of Eq. \ref{AsGa10} vanish so that%
\begin{equation}
J[\psi\left(  0\right)  ]=\frac{\gamma_{c}}{\hbar^{2}}\left[  \psi_{0}^{\ast
}\,\widehat{p}\,\psi_{0}+\psi_{0}\,\left(  \widehat{p}\,\psi_{0}\right)
^{\ast}\right]  =J^{f}[\psi\left(  0\right)  ]\label{AsGa12}%
\end{equation}
where $J^{f}[\psi\left(  0\right)  ]$ results from the application of the
free-electron current operator on $\psi\left(  0\right)  $. Thus, we obtain
another essential result: \textit{To first order, turning on the SOI does not
alter the value of the probability current.} Consequently, to solve the
problem we have only to show that $\psi$, given by Eq. \ref{AsGa2} and with
the conditions defined in Eq. \ref{AsGa11}, can match the boundary condition
expressed by Eq. \ref{AsGa6}. We obtain%
\begin{equation}
\gamma_{c}\left[  \alpha\widehat{p}+\hbar\left(  i\alpha\beta-\chi\right)
\right]  \psi_{0}=\gamma_{c}\left[  \alpha\widehat{p}+\hbar\left[
\alpha\left(  i\operatorname{Re}\beta+\chi\right)  -\chi\right]  \right]
\psi_{0}\text{ \ \ continuous.}\label{AsGa13}%
\end{equation}

The continuity of Eq. \ref{AsGa13} can always be ensured by taking $\alpha=1 $
and $\operatorname{Re}\beta=0$. Then, we calculate $\psi$ according to Eq.
\ref{AsGa2} and deduce the jump of $\psi$ at the interface%
\begin{equation}
\left[  \psi(0)\right]  _{0-}^{0+}=\left[  \frac{\chi}{\hbar(E-V)}\right]
_{0-}^{0+}\left(  \gamma_{c}\widehat{p}\psi_{0}\right)  =\mp\frac{1}{4\hbar
}\left[  \frac{\gamma}{\gamma_{c}^{2}}\right]  _{0-}^{0+}\left(  \gamma
_{c}\widehat{p}\psi_{0}\right)  .\label{AsGa14}%
\end{equation}

\section{Spin Current\label{Spin Cur}}

We have shown in Sec. \ref{sec GaAs} that Eq. \ref{860} provides a general and
symmetrized definition of the probability-current operator. Following the
conceptual scheme developed in Ref.
\onlinecite{Drouhin11}%
, we can define the spin currents in the up- and down-spin channels by taking
$\widehat{A}=\widehat{\pi}_{s}$, where $\widehat{\pi}_{s}$ is the orthogonal
projector on the spin basis ($s=\pm$). Then the SC current $\delta
\mathbf{J}_{\mathbf{u},j}\left[  \psi\right]  $, that arises from the
difference between the up-spin and the down-spin current, is obtained by
taking $\widehat{A}=\widehat{\sigma}_{\mathbf{u}}$, \ the Pauli operator along
the $\mathbf{u}$ direction defining the quantization axis. It is
straightforward to see that, as in Ref.
\onlinecite{Drouhin11}%
, the $j$-component of the spin-current operator is obtained from the
$j$-component of the probability-current operator after the substitution%
\begin{equation}
c_{j,l(1),...,l(n)}^{\prime}=\frac{1}{2}\{\sigma_{\mathbf{u}}%
,c_{j,l(1),...,l(n)}\}\text{.}\label{79bis}%
\end{equation}

\section{Conclusion}

We have proposed a systematic procedure to construct properly-symmetrized
current operators. We have extended the BenDaniel Duke
approach\cite{BenDaniel} to deal with heterostructures where SOI is included,
introducing generalized boundary conditions, which allow us to consider open
systems. We have shown that up to second order, usual matching conditions and
in particular the continuity of the envelope function at an interface, yield
solutions which comply with the conservation of the probability current. This
no longer holds as soon as a cubic term is included (D'yakonov-Perel' term).
We have illustrated our findings on the model case of a $[110]$-oriented GaAs
barrier, which had already been addressed in Ref.
\onlinecite{Hoai09}%
. We recover and generalize the results of Ref.
\onlinecite{Hoai09}
in a more simple and direct treatment. The tools we have developed can be
applied to the holes in the valence bands or to the electrons in the
conduction band so that they should be important for semiconductor-based spintronics.%

\begin{acknowledgements}%

We are deeply indebted to Travis Wade for a careful reading of the manuscript.%

\end{acknowledgements}%
%

\appendix

\section{Symmetry properties of current operators}

In Sec. \ref{sec GaAs}, Eq. \ref{8}, we derived the local form of the
Ehrenfest theorem for a general operator $\widehat{A}$ and deduced the
expression of the associated current $\mathbf{J}_{A}$. First, consider the
case where $\widehat{A}=\widehat{I}$, where $\widehat{I}$ is the identity and
the quadratic Hamiltonian $\widehat{\mathbf{p}}^{2}/2m$. We rewrite Eq.
\ref{8} as%
\begin{equation}
\frac{\partial}{\partial t}\left\vert \psi\right\vert ^{2}=-\mathbf{\nabla
}\cdot\operatorname{Re}\left(  \psi^{\dagger}\frac{\mathbf{\widehat{p}}}%
{m}\psi\right)  =-\mathbf{\nabla}\cdot\mathbf{J}\left[  \psi\right] \label{9}%
\end{equation}

We recover the usual expression for the free-electron probability current%
\begin{equation}
\mathbf{J}\left[  \psi\right]  =\operatorname{Re}\left(  \psi^{\dagger}%
\frac{\mathbf{\widehat{p}}}{m}\psi\right)  \text{.}\label{10}%
\end{equation}

Note that:%
\begin{equation}
\frac{\partial}{\partial t}\left\vert \psi\right\vert ^{2}=\frac{1}{i\hbar
}\left[  \left(  \psi^{\dagger}\frac{\mathbf{\widehat{p}}^{2}}{2m}\psi\right)
-\left(  \psi^{\dagger}\frac{\mathbf{\widehat{p}}^{2}}{2m}\psi\right)  ^{\ast
}\right]  =\frac{1}{i\hbar}\left[  \left(  \psi^{\dagger}\frac
{\mathbf{\widehat{p}}^{2}}{2m}\psi\right)  -\left(  \widehat{K}_{0}%
\psi\right)  ^{\dagger}\frac{\mathbf{\widehat{p}}^{2}}{2m}\left(  \widehat
{K}_{0}\psi\right)  \right] \label{11}%
\end{equation}
where $\widehat{K}_{0}$ is the time-reversal Kramers operator for a spinless
particle, which consists of taking the complex conjugate in the $\mathbf{r}%
$-representation. Let us check the expression of the current operators we
defined under time inversion symmetry. For this purpose we consider the term%
\begin{equation}
-2i\hbar\mathbf{\nabla\cdot J}_{A}=2i\operatorname{Im}\left(  \psi^{\dagger
}\left\{  \widehat{A},\widehat{H}\right\}  \psi\right)  =\left[  \psi
^{\dagger}\,\widehat{A}\,\widehat{H}\,\psi-\left(  \psi^{\dagger}\,\widehat
{A}\,\widehat{H}\,\psi\right)  ^{\ast}\right]  +\left[  \psi^{\dagger
}\,\widehat{H}\,\widehat{A}\,\psi-\left(  \psi^{\dagger}\,\widehat
{H}\,\widehat{A}\,\psi\right)  ^{\ast}\right]  \text{.}\label{17}%
\end{equation}

First, look at the term $\psi^{\dagger}\,\widehat{A}\,\widehat{H}\,\psi$%
\begin{align}
\left(  \widehat{K}\psi\left\vert \widehat{A}\,\widehat{H}\,\widehat{K}%
\,\psi\right.  \right)   &  =\left(  \widehat{K}_{0}\psi\left\vert \widehat
{R}^{\dagger}\,\widehat{A}\,\widehat{H}\,\widehat{K}\,\psi\right.  \right)
=\left(  \widehat{K}_{0}\psi\left\vert \widehat{R}^{\dagger}\,\widehat
{A}\,\widehat{K}\,\widehat{H}\,\psi\right.  \right)  =-\varepsilon_{A}\left(
\widehat{K}_{0}\psi\left\vert \widehat{R}^{\dagger}\,\widehat{K}\,\widehat
{A}\,\widehat{H}\,\psi\right.  \right) \nonumber\\
&  =-\varepsilon_{A}\left(  \widehat{K}_{0}\psi\left\vert \widehat{K}%
_{0}\,\widehat{A}\,\widehat{H}\,\psi\right.  \right)  =-\varepsilon_{A}\left(
\psi\left\vert \widehat{A}\,\widehat{H}\,\psi\right.  \right)  ^{\ast
}\label{18}%
\end{align}

Here, $\widehat{K}=\widehat{R}\,\widehat{K}_{0}$ is the Kramers operator for a
particle with spin $1/2$, $\widehat{R}=-i\sigma_{y}$ $\left(  \widehat
{R}^{\dag}=\widehat{R}^{-1}\right)  $, and $\varepsilon_{A}=\pm1$ depending
wether $\widehat{A}$ verifies\cite{Messiah}%
\begin{equation}
\widehat{K}\,\widehat{A}\,\widehat{K}=\varepsilon_{A}\,\widehat{A}%
\quad\text{or}\quad\widehat{R}^{\dagger}\,\widehat{A}\,\widehat{R}%
=\varepsilon_{A}\,\widehat{A}^{\ast}\text{.}\label{18 a}%
\end{equation}

Similarly, for the term $\psi^{\dagger}\,\widehat{H}\,\widehat{A}\,\psi$%
\begin{align}
\left(  \widehat{K}\psi\left\vert \widehat{H}\,\widehat{A}\,\widehat{K}%
\,\psi\right.  \right)   &  =-\varepsilon_{A}\left(  \widehat{K}_{0}%
\psi\left\vert \widehat{R}^{\dagger}\,\widehat{K}\,\widehat{H}\,\widehat
{A}\,\psi\right.  \right) \nonumber\\
&  =-\varepsilon_{A}\left(  \widehat{K}_{0}\psi\left\vert \widehat{K}%
_{0}\,\widehat{H}\,\widehat{A}\,\psi\right.  \right)  =-\varepsilon_{A}\left(
\psi\left\vert \widehat{H}\,\widehat{A}\,\psi\right.  \right)  ^{\ast}%
\text{.}\label{21}%
\end{align}

Thus, we obtain%
\begin{equation}
2i\operatorname{Im}\left(  \psi^{\dagger}\left\{  \widehat{A},\widehat
{H}\right\}  \psi\right)  =\psi^{\dagger}\left\{  \widehat{A},\widehat
{H}\right\}  \psi+\varepsilon_{A}\left(  \widehat{K}\psi\right)  ^{\dagger
}\left\{  \widehat{A},\widehat{H}\right\}  \left(  \widehat{K}\psi\right)
\text{.}\label{24}%
\end{equation}

We conclude that the general expression for the current of $\widehat{A}$ is%
\begin{equation}
\mathbf{\nabla}\cdot\mathbf{J}_{A}=-\frac{1}{2i\hbar}\left[  \psi^{\dagger
}\left\{  \widehat{A},\widehat{H}\right\}  \psi+\varepsilon_{A}\left(
\widehat{K}\psi\right)  ^{\dagger}\left\{  \widehat{A},\widehat{H}\right\}
\left(  \widehat{K}\psi\right)  \right]  \text{.}\label{25}%
\end{equation}

\section{Complete derivation of the current operator $\mathbf{\hat{J}}%
$\label{AppCurr}}

We are interested in finding the form of the current operator $\widehat
{\mathbf{J}}=\left(  \widehat{J}_{x},\ \widehat{J}_{y},\ \widehat{J}%
_{z}\right)  $ for an Hamiltonian $\widehat{H}^{\left(  n\right)  }+V\left(
\mathbf{r}\right)  $ - the current operator being $\widehat{\mathbf{J}%
}^{\left(  n\right)  }$ - and more generally for the Hamiltonian $\widehat
{H}_{eff}=\widehat{H}_{\mathbf{p}}+V\left(  \mathbf{r}\right)  =\sum
_{n}\widehat{H}^{\left(  n\right)  }+V\left(  \mathbf{r}\right)  $ (Eqs.
\ref{a 37}-\ref{37}) - the current operator being $\widehat{\mathbf{J}}$. For
an Hamiltonian $\widehat{\mathbf{p}}^{2}/2m+V\left(  \mathbf{r}\right)  $, it
is known\cite{MessiahJ} that the $j^{th}$ component of the current operator
($j=x,\ y,\ $\ or $z$) at the point $\mathbf{r}_{0}$ is of the shape
$\widehat{J}_{j}^{\left(  2\right)  }\left(  \mathbf{r}_{0}\right)  =\left(
1/2m\right)  \left[  \delta_{\mathbf{r}_{0}}\,\widehat{p}_{j}\,+\widehat
{p}_{j}\,\delta_{\mathbf{r}_{0}}\right]  $; With the notation of Eqs.
\ref{a 37}-\ref{37}, $\widehat{H}^{\left(  2\right)  }=\sum_{\overset
{l(k)\in\left\{  x,y,z\right\}  }{k=1,2}}c_{l\left(  1\right)  ,l\left(
2\right)  }\widehat{p}_{l\left(  1\right)  }\widehat{p}_{l\left(  2\right)  }%
$, $\widehat{J}_{j}^{\left(  2\right)  }\left(  \mathbf{r}_{0}\right)
=\sum_{l\left(  1\right)  =\left\{  x,\,y,\,z\right\}  }c_{j,l\left(
1\right)  }\left[  \delta_{\mathbf{r}_{0}}\,\widehat{p}_{l\left(  1\right)
}+\widehat{p}_{l\left(  1\right)  }\,\delta_{\mathbf{r}_{0}}\right]  $,
$c_{l\left(  1\right)  ,l\left(  2\right)  }=\left(  1/2m\right)
\delta_{l\left(  1\right)  ,l\left(  2\right)  }$. The aim of this appendix is
to show that, for an Hamiltonian $H^{\left(  n\right)  }+V\left(
\mathbf{r}\right)  $, the following form of the $j^{th}$ component of the
probability current operator
\begin{multline}
\hat{J}_{j}^{\left(  n\right)  }(\mathbf{r_{0}})=\underset{\overset
{l(k)\in\left\{  x,y,z\right\}  }{k=1,..,n-1}}{\sum}c_{j,l(1),...,l\left(
n-1\right)  }\left[  \delta_{\mathbf{r}_{0}}\widehat{p}_{l(1)}\widehat
{p}_{l\left(  2\right)  }...\widehat{p}_{l(n-1)}\right. \\
\left.  +\widehat{p}_{l(1)}\delta_{\mathbf{r}_{0}}\widehat{p}_{l\left(
2\right)  }...\widehat{p}_{l(n-1)}+...+\widehat{p}_{l(1)}\widehat{p}_{l\left(
2\right)  }...\widehat{p}_{l(n-1)}\delta_{\mathbf{r}_{0}}\right] \label{45}%
\end{multline}
gives back Eq. \ref{15}. The Dirac distribution interacts with the mixed
powers of the current operator so that the symmetrization procedure used in
the construction of $\hat{J}_{j}^{\left(  n\right)  }(\mathbf{r_{0}})$
provides $({n-2})$ further summations with respect to $\widehat{J}%
_{j}^{\left(  2\right)  }\left(  \mathbf{r}_{0}\right)  $. The two definitions
coincide only up to $n=2$. The extra terms are crucial in order to satisfy the
continuity equation. We evaluate every term over a generic state $\psi$; for
example the second term is of the shape%
\begin{align}
\left\langle \psi\left\vert \widehat{p}_{l(1)}\delta_{\mathbf{r}_{0}}%
\widehat{p}_{l\left(  2\right)  }...\widehat{p}_{l(n-1)}\right\vert
\psi\right\rangle  &  =\int\text{d}^{3}r\ \psi^{\ast}\widehat{p}_{l(1)}%
\delta_{\mathbf{r}_{0}}\widehat{p}_{l\left(  2\right)  }...\widehat
{p}_{l(n-1)}\psi\nonumber\\
&  =\int\text{d}^{3}r\ \left(  \widehat{p}_{l(1)}\psi\right)  ^{\dag}%
\delta_{\mathbf{r}_{0}}\widehat{p}_{l\left(  2\right)  }...\widehat
{p}_{l(n-1)}\psi\nonumber\\
&  =\left[  \widehat{p}_{l(1)}\psi\left(  \mathbf{r}_{0}\right)  \right]
^{\dag}\widehat{p}_{l\left(  2\right)  }...\widehat{p}_{l\left(  n-1\right)
}\psi\left(  \mathbf{r}_{0}\right)  \text{.}\label{46}%
\end{align}

Then the $j^{th}$ Cartesian component of probability current for a generic
state $J_{j}\left[  \psi\right]  $ can be written as:%
\begin{multline}
J_{j}^{\left(  n\right)  }\left[  \psi\right]  =\underset{\overset
{l(k)\in\left\{  x,y,z\right\}  }{k=1,..,n-1}}{\left\langle \psi\left\vert
\hat{J}_{j}^{\left(  n\right)  }(\mathbf{r_{0}})\right\vert \psi\right\rangle
=\sum c_{j,l(1),...,l\left(  n\right)  }}\left[  \psi^{\dagger}\widehat
{p}_{l(1)}...\widehat{p}_{l(n-1)}\psi+...\right. \\
\left.  +\left(  \widehat{p}_{l(1)}...\widehat{p}_{l(k-1)}\psi\right)
^{\dagger}\widehat{p}_{l(k)}...\widehat{p}_{l(n-1)}\psi+...+\left(
\widehat{p}_{l(1)}...\widehat{p}_{l(n-1)}\psi\right)  ^{\dagger}\psi\right]
\label{47}%
\end{multline}
where $\psi=\psi\left(  \mathbf{r}_{0}\right)  $. From Eq. \ref{47}, we can
find the generic divergence term related to the derivative with respect to
$\widehat{p}_{j}$:%
\begin{multline}
\widehat{p}_{j}J_{j}^{\left(  n\right)  }\left[  \psi\right]  =\underset
{\overset{l(k)\in\left\{  x,y,z\right\}  }{k=1,..n-1}}{\sum}%
c_{j,l(1),...,l\left(  n\right)  }\left[  \psi^{\dagger}\widehat{p}%
_{j}\widehat{p}_{l(1)}...\widehat{p}_{l(n-1)}\psi-\left(  \widehat{p}_{j}%
\psi\right)  ^{\dagger}\widehat{p}_{l(1)}...\widehat{p}_{l(n-1)}\psi\right. \\
+\left(  \widehat{p}_{l(1)}...\widehat{p}_{l(k-1)}\psi\right)  ^{\dagger
}\widehat{p}_{j}\widehat{p}_{l(k)}...\widehat{p}_{l(n-1)}\psi-\left(
\widehat{p}_{j}\widehat{p}_{l(1)}...\widehat{p}_{l(k-1)}\psi\right)
^{\dagger}\widehat{p}_{l(k)}...\widehat{p}_{l(n-1)}\psi\\
+\left(  \widehat{p}_{l(1)}...\widehat{p}_{l(k)}\psi\right)  ^{\dagger
}\widehat{p}_{j}\widehat{p}_{l(k+1)}..\widehat{p}_{l(n-1)}\psi-\left(
\widehat{p}_{j}\widehat{p}_{l(1)}...\widehat{p}_{l(k)}\psi\right)  ^{\dagger
}\widehat{p}_{l(k+1)}...\widehat{p}_{l(n-1)}\psi\\
\left.  +...+\left(  \widehat{p}_{l(1)}...\widehat{p}_{l(n-1)}\psi\right)
^{\dagger}\widehat{p}_{j}\psi-\left(  \widehat{p}_{j}\widehat{p}%
_{l(1)}...\widehat{p}_{l(n-1)}\psi\right)  ^{\dagger}\psi\right]
\text{.}\label{48}%
\end{multline}

In Eq. \ref{48} all the terms that have the same order in $k$ (two consecutive
terms but the first one and the last one) vanish after summation over $j$:%
\begin{multline}
\sum_{j=\left\{  x,y,z\right\}  }\sum_{\overset{l(k)\in\left\{  x,y,z\right\}
}{k=1,..n-1}}c_{j,l(1),}...,c_{l(n)}\,\left[  -\left(  \widehat{p}_{j}%
\widehat{p}_{l(1)}...\widehat{p}_{l(k-1)}\psi\right)  ^{\dagger}\widehat
{p}_{l(k)}...\widehat{p}_{l(n-1)}\psi\right. \\
\left.  +\left(  \widehat{p}_{l(1)}...\widehat{p}_{l(k)}\psi\right)
^{\dagger}\widehat{p}_{j}\widehat{p}_{l(k+1)}..\widehat{p}_{l(n-1)}%
\psi\right]  =0\label{48 bis}%
\end{multline}

Then the only terms still remaining in the summation are:%
\begin{align}
\sum_{j=\left\{  x,y,z\right\}  }\widehat{p}_{j}J_{j}^{\left(  n\right)
}\left[  \psi\right]   &  =\widehat{\mathbf{p}}\cdot\mathbf{J}^{\left(
n\right)  }\left[  \psi\right] \nonumber\\
&  =\sum_{j=\left\{  x,y,z\right\}  }\underset{\overset{l(k)\in\left\{
x,y,z\right\}  }{k=1,..,n-1}}{\sum}c_{j,l(1),}...,c_{l(n)}\left[
\psi^{\dagger}\widehat{p}_{j}\widehat{p}_{l(1)}...\widehat{p}_{l(n-1)}%
\psi-\left(  \widehat{p}_{j}\widehat{p}_{l(1)}...\widehat{p}_{l(n-1)}%
\psi\right)  ^{\dagger}\psi\right] \nonumber\\
&  =\sum_{j=\left\{  x,y,z\right\}  }\underset{\overset{l(k)\in\left\{
x,y,z\right\}  }{k=1,..,n-1}}{\sum}2i\ c_{j,l(1),}...,c_{l(n)}%
\operatorname{Im}\psi^{\dagger}\widehat{p}_{j}\widehat{p}_{l(1)}...\widehat
{p}_{l(n-1)}\psi\text{.}\label{a 49}%
\end{align}

Now $\mathbf{\nabla}\cdot\mathbf{J}^{\left(  n\right)  }\left[  \psi\right]
=\left(  i/\hbar\right)  \widehat{\mathbf{p}}\cdot\mathbf{J}^{\left(
n\right)  }\left[  \psi\right]  $ and Eq. \ref{49} results in a collection of
pure imaginary terms and the final expression for the divergence of the
probability current reads:%
\begin{equation}
\mathbf{\nabla}\cdot\mathbf{J}^{\left(  n\right)  }\left[  \psi\right]
=-\frac{2}{\hbar}\;\text{Im}\sum_{j=\left\{  x,y,z\right\}  }\;\;\underset
{\overset{l(k)\in\left\{  x,y,z\right\}  }{k=1,..,n-1}}{\sum}%
c_{j,l(1),...,l\left(  n\right)  }\left(  \psi\left\vert \widehat{p}%
_{j}\widehat{p}_{l(1)}...\widehat{p}_{l(n-1)}\right\vert \psi\right)
\text{.}\label{49}%
\end{equation}

Eventually%
\[
\mathbf{\nabla}\cdot\mathbf{J}\left[  \psi\right]  =\sum_{n}\mathbf{\nabla
}\cdot\mathbf{J}^{\left(  n\right)  }\left[  \psi\right]  .
\]


\begin{thebibliography}{99}
\bibitem {JeanEric}A thermodynamic model only based on the continuity equation
Eq. \ref{3} cannot take into account the exchange of energy $e$ with the
environment introduced by the current generator and the spin-orbit
interaction. Indeed, the consequence of the Gibbs equation, d$e/$d$t=T\partial
s/\partial t+\mu\partial\rho/\partial t$ (where $s$ is the entropy, $T$ the
temperature, and $\mu$ is the chemical potential), is that the continuity
equation of the density of carriers $\rho$ is necessarily determined by the
power dissipated at the boundaries. At zero temperature, the source term is
equal to the power dissipated divided by the chemical potential $\partial
\rho/\partial t=(1/\mu)$d$e/$d$t$. This is the reason why the decomposition
between the divergence term and the source term $G$ cannot be unique without
specifying the boundary conditions.
\end{thebibliography}
\end{document}